# A Conceptual E-Governance Framework for Improving Child Immunization Process in India

Puneet Kumar
Assistant Professor
Deptt. Of Physical & Computational Science
Mody Institute of Technology & Science,
Lakshmangarh, Sikar, Rajasthan, India

Dharminder Kumar
Professor & Chairman
Department of Computer Science & Engineering
Guru Jambheshwar University of Science &
Technology, Hisar, Haryana, India

## ABSTRACT

India is country having high population and great variations in the educational level, economic conditions, population densities, cultures and awareness levels. Due to these variations the immunization process is not so much successful as per expectations of the state and central governments. In some zones the significant amount of vaccines are wasted whereas some are running out of vaccines. One of the reasons for such an imbalance is improper quantity estimation of vaccines in a particular zone. Further a huge amount of liquidity will be wasted in the form of vaccines. If we inculcate ICT (Information and Communication Technology) in the process of immunization then the problem can be rectified to some extent and hence we are proposing a conceptual model using ICT to improve the process of vaccination.

## 1. INTRODUCTION
In the field of e-governance, India is having many islands of success but still there many areas which are still unexplored or explored in an inadequate manner. Child immunization is one which is partially explored.  Although India is having IT (Information Technology) infrastructure up to village level but despite of that vaccination process is still without IT supplement. The immunization process is not fully computerized and still papers are being used for maintaining various types of records. Since ICT is not being used in this area therefore the speed of communication is not fast which results into various types of problems like demand and supply of drugs in particular areas, underflow or overflow of drugs etc. [1]. The current system of vaccination is inflexible because the vaccination card is mandatory at the time of child vaccination. If someone loses the vaccination card then the immunization records related to a particular child can't be fetched instantly. The Govt. of India is spending a significant amount on vaccination process but still we are able to hit only a target of 80 percent [2]. In this paper we have tried to propose a conceptual framework which can improve the process of vaccination as well as management of vaccines to some extent. The devised conceptual frame work can automate the child immunization process to a great extent and further, it will also help state and central governments in policy implementation and decision making process.

## 2. REVIEW OF LITERATURE
E-governance endures the use of Information and Communication Technology (ICT) to promote more efficient and effective governance. It facilitates access to government services, access to public information in a better manner and makes the system more transparent and accountable [3]. India is the second largest country having 1.2 billion populations and it's a big challenge for the government to deliver e-services to such a huge populace. [4] To conquer the challenges, government of India has formulated the National e-Governance Plan (NeGP) comprises of various mission mode projects for facilitating the people with e-services in almost every dimension of governance. [5] For availing various types of e-services, individual monitoring, better planning and preventing malpractices in various aspects, a unique identification number has been issued to every citizen of India known as 'AADHAR' number. It resembles with the identity of every individual and the identity can be verified online. [6] The government of Rajasthan has initiated an e-governance project in the area of health care. It tracks and monitors every pregnant woman and infant child for imparting better health services. The central objective behind such a project is to reduce the child and maternal mortality rate by providing proper and hygienic delivery of the child and proper immunization to the infant child.[7] The reports reveals that the situation of India is pathetic regarding immunization among children. As per the 2001 Census, 3.45 crore persons were residing in towns and cities of Uttar Pradesh and the data discloses that only 23% children aged 12-23 months receiving complete vaccination. Only 15.3% children are completely immunized [8]. More than half of the children (51.9%) were un-immunized. One third of the mothers (34.7%) had no idea about the place of immunization. Among the children who were either partially or fully immunized, 77.9% had received immunization from the government health facility. The most common reason behind this was the lack of information (77.2%). The dropout rate for complete immunization was found to be 32.7% [9].India has one of the lowest routine immunization rates in the world [10]. Estimates from the Indian National Family Health Survey indicate that only 43.5% of children age 12-23 months were fully vaccinated and 5% had received no vaccinations at all [11]. India is also undergoing with a severe wastage of vaccines. A survey was conducted by UNICEF in 2010 in five states viz Uttar Pradesh, Assam, Maharashtra, Tamil Nadu, Himachal Pradesh and the facts revealed by UNICEF are surprising i.e. according to survey the average wastage rate of BCG vaccine was 61%, 47% for OPV, 35% for measles, 34% for TT, 33% for Hepatitis B and 27% for DPT [12].

## 3. CASE STUDY
### 3.1 PCTNS-Rajasthan
The pregnancy, child tracking & health services management system (PCTNS) is an award winning [13] G2C (Government to Citizen) application which is developed by National Informatics Centre and implemented in Rajasthan state. It's a landmark project in the area of health care with the vision of providing better health services to rural and urban masses especially towards economically weaker section of society. The main objective of the project is to reduce maternal mortality rate and child mortality rate [14].

*3.1.1 Features*
PCTNS performs online tracking for every pregnant woman who has registration in any PHC (Primary Health Centre). The tracking will be done for; providing antenatal health services, delivery services, every single abortion case and maternal death. The system is online and having centralized database which ensures the availability of historical and current data all the time. It also performs online tracking of every single live birth along with associated attributes like birth's weight, hemoglobin, immunization and vitamin A doses. It also monitors for area/district wise sex ratio and drop outs and left outs from immunization.

*3.1.2 Limitations*
3.1.2.1 The system is not nationwide and hence confined to a single state, so it can't be used for centralized planning for whole country.

3.1.2.2 The 'Aadhar' number is not being used in by the system which ensures one's identity.

3.1.2.3 The main focus is towards imparting health services to pregnant woman.

3.1.2.4 Online tracking can't be performed if a mother or child is relocated outside the state.

3.1.2.5 Not able to produce various types of certificates like birth certificate, death certificate (incase mother or child dies) etc.

3.1.2.6 The tracking can be performed only when a woman registers her selves in a government health center.

3.1.2.7 It's an in-house project developed by NIC i.e. private partners are not involved.

## 4. PROPOSED FRAMEWORK
As per Unique Identification Authority of India, every citizen will be having a unique identification number i.e. Aadhar which uniquely identifies a person across the country. Since there no minimum age limit for Aadhar [15] therefore infants can be registered on the behalf of their parents or guardians [16]. Every health centre either government or private should be equipped with a small UID generation unit or child registration unit.

The figure-1 demonstrates the framework for initial vaccination and child registration process and the steps are given as under:

Step1: The private/ Government health centre will invoke the infant registration module and parent or guardian or owner of the orphanage (in case the child is an orphan) will provide their Aadhar number provided by government of India for the initiation of infant registration process.

Step2: The new UID registration module will be accessed from UID server and request for new UID will be made.

Step3: The data provided with the request will be verified and a new UID will be created.

Step4: Some of the required attributes related to newly registered infants will be replicated in the central immunization database like child's name, guardian's name, guardian's contact number, date of birth etc.

Step5: The UID for the infant will be gathered.

Step6: Thereafter the birth certificate will be issued to the guardian comprising of all birth details along with UID.

Step7: The initial vaccination registration module will be invoked from immunization server and request for child registration for immunization will be made. After verification all the details related to the infant will be entered along with the entries related to the vaccines given to the infant.

Step8: The child & vaccine registration module will be fetched form the immunization server and request for initial vaccination will be made.

Step9: The data will be verified and registration of child as well as vaccines given to child will be made.

Step10: A message will be sent at the parent/ guardian's mobile containing the information about the vaccines given to the child and next date for vaccination.

Step11: A report comprises of the details related to the registered infants can be sent to municipal corporations for their use.

The figure 2 demonstrates the framework for the whole immunization process after initial vaccination and the steps are given as under:

Step1: The revaccination module will be invoked with given child UID

Step2: The child & vaccine registration module will be requested for revaccination entry after verification.

Step3: The details will be verified and revaccination entries will be made in the records.

Step4: A message will be sent to parent/ guardian at their mobile phones containing information related to currently given vaccines as well as the date for next vaccination.

### 4.1 Comparative analysis between proposed framework and PCTNS
The table given below indicates the contrast between existing e-health service and proposed framework.

**Table 1: Comparison between PCTNS & proposed model**

| Feature | PCTNS | Proposed |
|---|---|---|
| Can be implemented at National level? | NO | YES |
| Primary Key | Local | UID number i.e. 'Aadhar' |
| Accessibility | Only from government health centers | private and government health centers |
| Focus | Pregnant women and infant child | Exclusively children |
| Generation of birth/ death certificates | NO | Birth certificate |
| Can help in planning at national level? | NO | YES |
| Connectivity with municipal corporation | NO | YES |
| Can implemented through PPP (Public Private Partnership)? | Already implemented | YES |

It's evident from above that the proposed framework will enhance the flow of information and provide more mobility to the citizen regarding vaccination.

## 4.2  Benefits in adopting proposed model

The proposed model is definitely an enhancement in services provided by government to citizens. It will benefit both i.e. government as well as citizen. The major advantages can be enlisted as:

a. It will provide more flexibility to the parent or guardian i.e. a parent/ guardian may go to any health centre irrespective of same city/town/village, same district or same state.

b. Since all the information is online and UID card will be sufficient for tracking all the details therefore no need to issue vaccination card. It will save a huge amount of papers.

c. The government can access the data very easily because it is centralized therefore it can help a lot in policy making and analyzing current scenario like infant mortality information, success percentage of immunization, drugs requirements in a particular area or a particular state or for whole country etc.

d. It will also help in demographic analysis for the particular areas or districts or states.

e. It will also make the flow of information very fast so that quick and accurate decisions can be taken in a real time.

## 5. CONCLUSION

The proposed model is a conceptual model and, if implemented, it will be a valuable addition in the services provided by government to citizens. The implementation of the proposed model doesn't require overwhelming costs because in India we are already having functional ICT (Information and Communication Technology) infrastructure at all the levels and as a part of National e-Governance Plan every village is about to be equipped with a CSC (Common Service Centre). Since the proposed model provides flexibility to the common man therefore it will certainly going to increase the percentage of immunization in whole country and it will be a successful step to make healthy India.

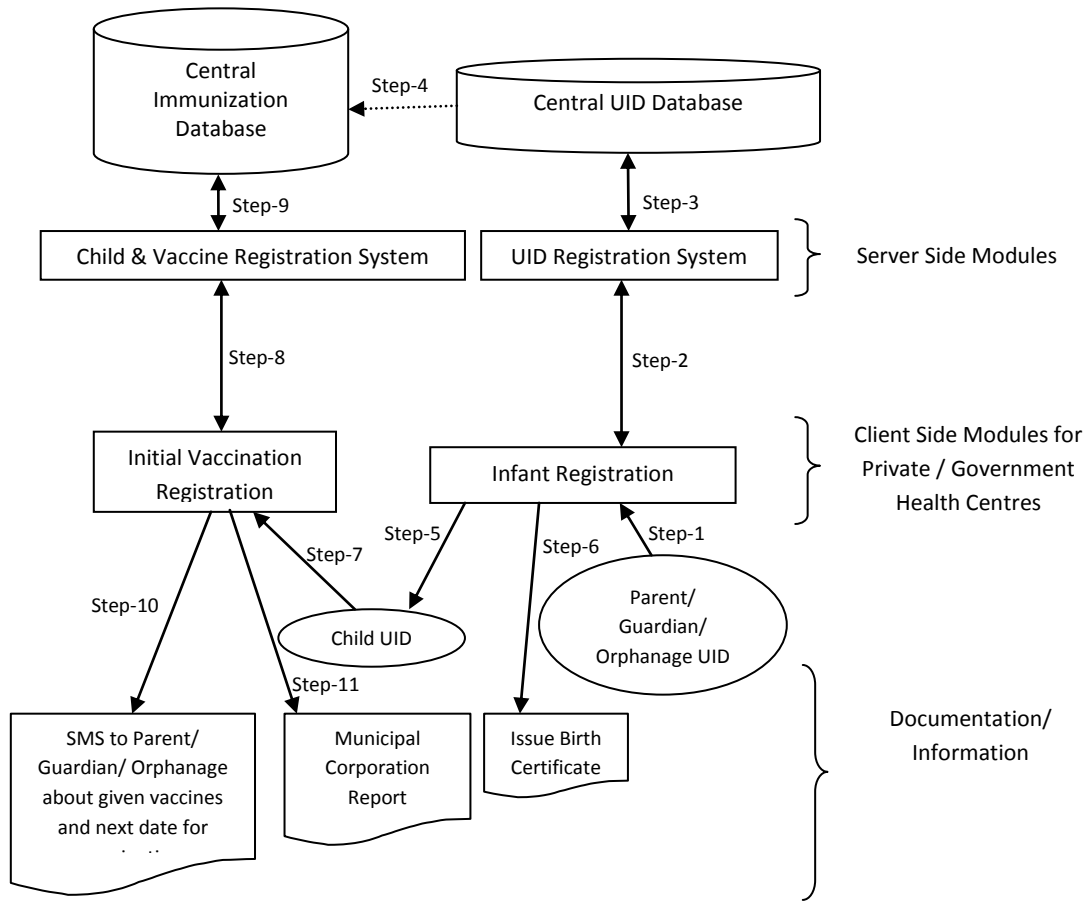

**Fig 1: Framework for initial vaccination and child registration process**

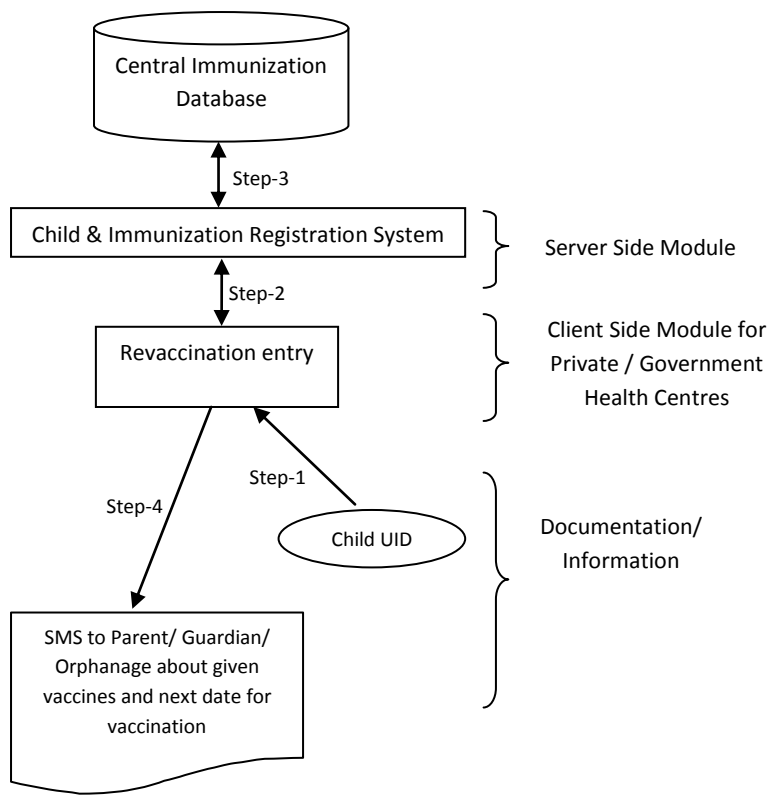

**Fig 2: Framework for the whole immunization process after initial vaccination**